\documentclass[aps,prl,twocolumn,notitlepage,superscriptaddress]{revtex4-1}

\usepackage{bm} \usepackage{graphicx} \usepackage{amsmath}
\usepackage{amsthm,stmaryrd}
\usepackage{amssymb} 
\usepackage{physics} 
\usepackage{txfonts}
\usepackage{color}
\usepackage{hyperref}
\usepackage{complexity}

\usepackage{cleveref}

\crefname{equation}{}{}

\newcommand{\comment}[1]{}

\newcommand{\idty}[1]{\mathbb{1}}
\newcommand{\ovsqrt}[1]{\frac{1}{\sqrt{2}}}

\usepackage{algpseudocode}

\newcommand{\N}{\mathbb{N}}

\newcommand{\Ecal}{\mathcal{E}}

\usepackage{xparse}
\NewDocumentCommand\Exp{g}{%
    \ensuremath{\mathbb{E}\IfNoValueTF{#1}{}{\left[#1\right]} }%
}

\newcommand{\Ex}{\operatorname*{\mathbb{E}}\nolimits}

\DeclareMathOperator*{\argmin}{arg\,min}

\renewcommand{\paragraph}[1]{~\newline\noindent {\itshape #1.}}

\begin{document}

\title{Statistical Limits of Supervised Quantum Learning}

\author{Carlo Ciliberto}
\affiliation{	Department of Electrical and Electronic Engineering, Imperial College London, London SW7 2BT, United Kingdom}

\author{Andrea Rocchetto}
\affiliation{Department of Computer Science, University of Texas at Austin, Austin, TX 78712, USA}
\affiliation{Kavli Institute for Theoretical Physics, University of California, Santa Barbara, CA 93106, USA}

\author{Alessandro Rudi}
\affiliation{INRIA - Sierra project team, Paris 75012, France}

\author{Leonard Wossnig}
\affiliation{Department of Computer Science, University College London, London WC1E 6EA, United Kingdom}
\affiliation{Rahko Limited, N4 3JP London, United Kingdom}  

\begin{abstract}
Within the framework of statistical learning theory it is possible to bound the minimum number of samples required by a learner to reach a target accuracy. 
We show that if the bound on the accuracy is taken into account, quantum machine learning algorithms for supervised learning---for which statistical guarantees are available---cannot achieve polylogarithmic runtimes in the input dimension. We conclude that, when no further assumptions on the problem are made, quantum machine learning algorithms for supervised learning can have at most polynomial speedups over efficient classical algorithms, even in cases where quantum access to the data is naturally available.
\end{abstract}

\maketitle

A wide class of quantum algorithms for supervised learning problems (where the goal is to infer a mapping given examples of an input-output relation) exploit fast quantum linear algebra subroutines to achieve runtimes that are exponentially faster than their classical counterparts~\cite{biamonte2017quantum, ciliberto2018quantum}.
Examples of these algorithms are quantum support vector machines~\cite{rebentrost2014quantum}, quantum linear regression~\cite{wiebe2012quantum,schuld2016prediction}, and quantum least squares~\cite{kerenidis2017quantum, chakraborty2018power}.

A careful analysis of these algorithms identified a number of caveats that limit their practical applicability such as the need for a strong form of quantum access to the input data, restrictions on structural properties of the data matrix (such as condition number or sparsity), and modes of access to the output~\cite{aaronson2015read}.
Furthermore, if one assumes that it is efficient to (classically) sample elements of the training data in a way proportional to their norm, then it is possible to show that classical algorithms  are only polynomially slower (albeit the scaling of the quantum algorithms can be considerably better)~\cite{tang2018quantum,chia2018quantum,chia2019quantum,gilyen2018quantum, chia2019samplingbased}.

In this paper we continue to investigate the limitations of quantum algorithms for supervised learning problems.
Our analysis focuses on the dependency on the size of the data set that is introduced when considering the statistical guarantees of the estimators.
The key elements of our work are a series of well known results in statistical learning theory that show how the accuracy parameter of a supervised learning problem scales inverse polynomially with the number of samples in the training set.
We leverage on these insights to show that quantum learning algorithms must have at least polynomial runtime in the dimension of the training data and therefore cannot achieve exponential speedups over classical polynomial time machine learning algorithms. 
We remark that our results do not rule out exponential advantages for learning problem where no efficient classical algorithms are known. 
In fact, in this regime, there exist learning problems for which quantum algorithms have a superopolynomial advantage~\cite{grilo2019learning, kanade2019learning}.

Our results are independent of the modes of access to the training data, that is, even if the data set is naturally stored in a quantum structure, quantum machine learning algorithms can have at most polynomial advantage over their classical variants.

Finally, we note that our results do not assume any prior knowledge on the function to be learned.
This allows us to make statements on virtually every possible learning algorithm, including neural networks.
Using stronger assumptions on the target function it is possible to improve the dependency of the accuracy in number of samples (consider the limit case where the function is known, here zero samples can determine the function with maximum accuracy).

\paragraph{Statistical Learning Theory}
The field of statistical learning theory investigates how to quantify the statistical resources required to solve a learning problem~\cite{shalev2014understanding}. In this work, we consider supervised learning settings where the goal is to find a model that fits well a set of input-output training examples but that, more importantly, guarantees good prediction performance on new observations. This latter property, also known as {\em generalisation capability} of the learned model, is a key aspect separating machine learning from the standard optimisation literature. Indeed, while data fitting is often approached as an optimisation problem in practice, the focus of machine learning is to design statistical estimators able to `fit' well future examples.

More formally, let $\rho$ be a distribution over $X \times Y$, with $X$ a so-called \textit{domain (or input) set} and $Y$ a \textit{label (or output) set}. The goal of supervised learning is to produce a hypothesis $f: X \rightarrow Y$ such that the \textit{expected risk} or \textit{expected error}
\begin{equation}\label{eq:expected-risk}
    \Ecal( f):= \Ex_{\rho} \left[ \ell(y, f(x))  \right]
\end{equation}
is small with respect to a suitable \textit{loss function} $\ell : Y \times Y \rightarrow \mathbb{R}$ measuring prediction errors. However, in practice, the target distribution $\rho$ is unknown and only accessible by means of a finite {\itshape training set}  $S_n = \{(x_i, y_i), i=1, \dots, n \}$ of i.i.d. points sampled from it. 

Depending on whether the label set $Y$ is dense or discrete the task is called \textit{regression} (dense) or \textit{classification} (discrete). Typical loss functions are the quadratic loss $\ell_{\mathrm{sq}}(f(x), y) = ( f(x) - y )^2$ over $Y=\mathbb{R}$ for regression and the $0-1$ loss $\ell_{0-1}(f(x), y) = \mathbf{1}_{f(x) \neq y}$ over $Y=\{-1,1\}$ for classification.

Different machine learning frameworks have different prescriptions on how to choose the hypothesis $f$.
The \textit{Empirical Risk Minimisation} (ERM) framework prescribes to choose a hypothesis that minimises the empirical risk
\begin{equation}\label{eq:erm}
    \inf_{f\in \mathcal{H}} \hat{\Ecal} (f), \qquad \hat{\Ecal}(f) = \frac{1}{n} \sum_{(x_i,y_i) \in T} \ell(y_i, f(x_i)),
\end{equation}
over a suitable {\itshape hypotheses space} $\mathcal{H}$. Under weak assumptions on $\mathcal{H}$ (for instance a bounded subset of a Hilbert space [16]). it is possible to guarantee the existence of a minimizer for \cref{eq:erm} that we denote $\hat f = \argmin_{f\in\mathcal{H}}\hat\Ecal(f)$.

The difference between risk and empirical risk is called {\itshape generalisation error} and plays a central role in statistical learning theory. Indeed, when \cref{eq:expected-risk} admits a minimizer in $\mathcal{H}$, we have
\begin{equation}
\label{eq:genbound}
\Ecal(\hat f) - \inf_{f\in \mathcal{H}} \Ecal (f) \,\leq 2\sup_{f\in\mathcal{H}}\left| \hat{\Ecal}(f) - \Ecal(  f) \right|.
\end{equation}
In other words, the {\itshape excess risk} incurred by the empirical risk minimizer is controlled by the worse generalisation error over $\mathcal{H}$. 
A fundamental result in statistical learning theory~\cite{vapnik1998, blumer1989learnability, shalev2014understanding}, often referred in the literature as the fundamental theorem of statistical learning, is that for every $n\in\N$, $\delta \in (0,1)$, and every distribution $\rho$, the following holds with probability larger than $1-\delta$
\begin{equation}
\label{eq:errorscaling}
\sup_{f\in\mathcal{H}}\left| \hat{\Ecal}(f) - \Ecal(  f) \right| \leq \Theta \left(  \sqrt{\frac{c\,(\mathcal{H})+\log(1/ \delta)}{n} }
    ~\right),
\end{equation}
where $c\,(\mathcal{H})$ is a measure of the complexity of $\mathcal{H}$ (such as the VC dimension, covering numbers, Rademacher complexity to name a few~ \cite{cucker2002mathematical,shalev2014understanding}). 
Intuitively, the dependency on $c(\mathcal{H})$ in \cref{eq:errorscaling} models the phenomenon known as \textit{overfitting} in which a large hypothesis space incurs in low training (empirical) error but performs poorly on the true risk.
This problem can be addressed with so-called \textit{regularisation techniques}, which essentially limit the expressive power of the learned estimator in order to avoid overfitting the training dataset.

Different regularisation strategies have been proposed in the literature (see~\cite{vapnik1998,bishop2006pattern,bauer2007regularization} for an introduction to the main ideas), and one of the well-established approaches which directly imposes constraints on the hypotheses class of candidate predictors is the Tikhonov regularisation.
Regularisation ideas have led to popular machine learning approaches which are widely used in practice, such as Regularised Least Squares~\cite{cucker2002mathematical}, Gaussian Process (GP) Regression and Classification \cite{rasmussen2006gaussian}, Logistic Regression~\cite{bishop2006pattern}, and Support Vector Machines (SVM)~\cite{vapnik1998}. 
All these algorithms can be studied within the framework of kernel methods~\cite{shawe2004kernel}.

From a computational perspective, these approaches compute a solution for the learning problem by optimising over the constraint objective, which typically consists of a sequence of standard linear algebra operations such as matrix multiplication and inversion. For most classical algorithms, such as GP or SVM, the computational time is therefore $\mathcal O(n^3)$, which is similar to the time it requires to invert a square matrix that has size equal to the number $n$ of examples in the training set. 
Notably this can be improved depending on the sparsity and the conditioning of the specific optimisation problem. 

To reduce the computational cost, instead of considering the optimisation problem as a separate process from the statistical one, more recent methods hinge on the intuition that reducing the computational burden of the learning algorithm can be interpreted as a form of regularisation on its own. 
For instance, {\em early stopping} approaches are now widely used in practice, and perform only a limited number of steps of an iterative optimisation algorithm, to avoid overfitting the training set. 
They thereby entail less operations, while provably maintaining the same generalisation error of approaches such as Tikhonov regularisation~\cite{bauer2007regularization}. 
More specifically, prototypical results (such as~\cite{bauer2007regularization}) show that the number of iterations required are of the order of $1/\lambda$ where $\lambda$ is the ideal regularisation parameter that one would use for ERM. Therefore, if in the worst case scenario $\lambda = O(1/\sqrt{n})$, early stopping would attain (up to constants) the same generalisation error of regularised ERM by performing only $\sqrt{n}$ iterations.

A different approach, also known as {\em divide and conquer}, is based on the idea of distributing portions of the training data onto separate machines, each solving a smaller learning problem, and then combining individual predictors into a joint one. 
This computation hence benefits from both the parallelisation and the reduced dimension of distributed datasets while similarly maintaining statistical guarantees~\cite{zhang2013divide}.

A third approach that has recently received significant attention from the machine learning community, along with the quantum computing community, is based on random sub-sampling, a form of dimensionality reduction. Depending on how such sampling is performed, different methods have been proposed, the most well-known being random features~\cite{rahimi2007random} and Nystr\"om approaches~\cite{smola2000sparse,williams2001using}. 
Here the computational advantage is clearly given by the smaller dimensionality of the hypotheses space, and it has also recently been shown that it is possible to obtain equivalent generalisation error to classical methods in these settings~\cite{rudi2015less}.

For all these methods, training times can be typically reduced from the $\mathcal O(N^3)$ of standard approaches to $\widetilde{ \mathcal{O}} (N^2)$ or $\widetilde{ \mathcal{O}} (Nz)$, where $z$ is the number of non-zero entries, while keeping the statistical performance of the learned estimator essentially unaltered.

\paragraph{Quantum learning algorithms}
Linear algebra subroutines are a central computational element of learning algorithms. 
A large class of quantum algorithms for supervised learning problems claim exponential speed-ups compared to classical algorithms
by making use of fast quantum linear algebra subroutines~\cite{wiebe2012quantum,schuld2016prediction,rebentrost2014quantum,wang2017quantum,kerenidis2017quantum,chakraborty2018power,zhang2018nonlinear}.
One widely used algorithm is the quantum linear system solver~\cite{harrow2009quantum} (also known as HHL after the three authors Harrow, Hassidim, and Lloyd). The algorithm takes as input a quantum encoding of the vector
$b \in \mathbb{R}^n$ and a $s$-sparse matrix $A \in \mathbb{R}^{n \times n}$, with $\norm{A} \leq 1$, and outputs an approximation $\ket{\tilde{w}}$ of the solution $\ket{w=A^{-1}b}$ of the linear system such that
\begin{equation}
\norm{\ket{\tilde{w}} - \ket{w}} \leq \gamma
\end{equation}
for an error parameter $\gamma > 0$.
The current best implementation of the algorithm runs in time~\cite{chakraborty2018power}
\begin{equation}
    O(\norm{A}_F \, \kappa\, \polylog (\kappa, n, 1/\gamma)),
\end{equation}
where $\norm{A}_F$ is the Frobenius norm of $A$ and $\kappa$ its condition number. 
Note that the HHL algorithm requires to access the data matrix $A \in \mathbb{R}^{n\times d}$ in $O(\polylog(nd))$ time.
All the quantum learning algorithms we discuss in this paper inherit this assumption.
Recently, it was proven that such strong oracular assumptions (when the data matrix is low-rank) also lead to
exponentially faster classical algorithms~\cite{tang2018quantum,chia2018quantum,gilyen2018quantum, chia2019samplingbased}.
We recommend~\cite{ciliberto2018quantum,aaronson2015read} for more detailed discussions of the limits of quantum learning algorithms based on fast linear algebra subroutines.

Before proceeding to the statistical analysis of quantum learning algorithms we review some quantum algorithms for the least squares problem.
These will serve as the main examples in our analysis.

\paragraph{Quantum least squares}
Least squares is an algorithm for minimising the empirical risk, with respect to the squared loss, for the hypothesis class of linear functions.
More specifically let $X = \mathbb{R}^d$ and $Y= \mathbb{R}$, and let $\mathcal{H} := \{ f : X \rightarrow Y \, \vert \, \exists \, w \in \mathbb{R}^d : f(x) = w^T x \}$ be the hypothesis class of linear functions. The empirical risk is
\begin{equation}
    \hat{\Ecal} (f) := \frac{1}{n} \sum_{i=1} ^n \left(w^T x_i -y_i \right)^2.
\end{equation}
We can minimise the empirical risk by setting its gradient to zero. Using $X:= \sum_i x_i x_i ^T$ and $b := \sum_i y_i x_i$ one can write a close form solution to the least squares problem as $w = X^{-1} b$.

Several quantum algorithms for least squares (or, more generally, linear regression problems) have been proposed~\cite{wiebe2012quantum,kerenidis2017quantum,chakraborty2018power,wang2017quantum,zhang2018nonlinear}. A common feature is that they use a fast quantum linear system algorithm to find a quantum encoding $\ket{w}$ of the solution vector $w = X^{-1} b$.
The fastest known algorithm in the class~\cite{chakraborty2018power}, which improves the dependency on the error from polynomial to logarithmic, solves the (regularized) least squares or linear regression problem in time 
\begin{equation}
O(\norm{X}_F \, \kappa \, \polylog(n,\kappa,1/\gamma)),
\end{equation}
where $\kappa$ is the condition number of $X$ and $\gamma>0$ is an approximation parameter. As for every other quantum algorithm discussed in this paper the quantum least squares solver requires a quantum-accessible data structure.
The dependency on the Frobenius norm implies that it is possible to obtain a speedup only when $X$ is low-rank (but non-sparse).
Due to approximation errors, the output of the algorithm is not $\ket{w}$ but a quantum state $\ket{\tilde{w}}$, such that $\norm{\ket{\tilde{w}}-\ket{w}} \leq \gamma$.

It is possible to get rid of the dependency on the Frobenius norm using the sample
based Hamiltonian simulation method~\cite{lloyd2014pca,kimmel2017hamiltonian}.
Leveraging this technique,~\cite{schuld2016prediction} proposed a least squares algorithm whose scaling does not depend on the Frobenius norm but requires a higher number copies (with respect to~\cite{chakraborty2018power}) of the input
density matrix.
Note that, because the algorithm in~\cite{schuld2016prediction} is posed in the query model, i.e. the computational complexity
is given in number of calls to the oracle which returns the data already encoded in form of a quantum state, it is not possible to make a direct comparison between the two algorithms.
The computational complexity of the algorithm given in~\cite{schuld2016prediction} is 
\begin{equation}
O(\kappa^2 \gamma^{-3} \polylog(n)), 
\end{equation}
and the dependency on the error is polynomial.


\paragraph{Quantum speed-ups and statistical bounds}
In this section we analyse the speed-up claims of quantum machine learning algorithms using the framework of statistical learning theory.
Our main point is that if one considers  the $\Theta(n^{-1/2})$ scaling of the generalisation error---see \cref{eq:errorscaling}---quantum learning algorithms cannot achieve polylogarithmic runtime in $n$.

The starting point of our discussion is the following standard error decomposition. 
Consider an hypothesis $f$.
We want to bound how far the generalisation error of $f$ is from the best possible generalisation error; this is known as the \textit{Bayes risk} and is indicated by $\Ecal^* := \inf_{f\in \mathcal{F}} \, \Ecal( f)$, where $\mathcal{F}$ denotes the set of all measurable functions $f:X\rightarrow Y$. 
We want to decompose this general error into different components and for this reason we introduce $\Ecal_{\mathcal{H}}:= \inf_{f\in \mathcal{H}} \, \Ecal( f)$, that is the best risk attainable by function in the hypothesis space $\mathcal{H}$. In order to simplify our discussion let us assume that $\Ecal_{\mathcal{H}}$ always admits a minimizer $f_{\mathcal{H}}\in\mathcal{H}$ (it is possible to levy this assumption using the theory of regularisation).
Recalling that $\hat{\Ecal}(\hat{f}) := \inf_{f \in \mathcal{H}} \hat{\Ecal}(f)$, we can decompose the total error as:
\begin{align}
\label{eq:totalerrordec}
    \Ecal( f) - \Ecal^* &= \underbrace{\Ecal( f) - \Ecal( \hat{f})}_{\textrm{Optimisation error}} ~+~ \underbrace{\Ecal( \hat{f}) - \Ecal_{\mathcal{H}}}_{\textrm{Estimation error}} ~+~ \underbrace{ \Ecal_{\mathcal{H}} - \Ecal^*}_{\textrm{Irreducible error}}  \\
     &= \xi + \Theta(1/\sqrt{n}) + \mu. 
\end{align}
The first term in \cref{eq:totalerrordec} is the \textit{optimisation error} and measures how good is the optimisation that generated $f$ with respect to the ideal minimisation of the empirical risk. 
This error is related to the approximation error of the algorithm. 
The second term is the \textit{estimation error} and models the error that we make by estimating the true risk using samples from the distribution $\rho$.
This is the generalisation bound we discussed in \cref{eq:errorscaling}.
The third term is the \textit{irreducible error} and measures how well the hypothesis space models the problem.
It is an irreducible source of error that we indicate with the letter $\mu$.
If the irreducible error is zero than we say that $\mathcal{H}$ is universal.
For simplicity, we assume throughout the paper that $\mu = 0$.

From the error decomposition in  \cref{eq:totalerrordec} we see that in order to have an algorithm with optimal statistical performance we must make sure that the optimisation error is not larger than the estimation error.
Therefore the optimisation error must scale at most as the best estimation error. If it does, we say that the optimisation error matches the bound of the estimation error.

In order to make the notion of matching the bound more concrete, let us consider again the case of least squares.
The closed form solution $w = X^{-1}b$ requires $O(n^3)$ time to be computed and attains essentially zero optimisation error.
Because the total error is dominated by the $1/\sqrt{n}$ term of the estimation error, one may wonder about the convenience of paying a cost of order $O(n^3)$ to achieve zero optimisation error.
A careful analysis shows that this is indeed not a convenient choice and it is possible to design algorithms that are less accurate but converge faster to estimators that, albeit not attaining zero optimisation error, achieve an error that matches the bound---this is the approach taken by early stopping, divide and conquer, and random sub-sampling methods.
For many quantum algorithms, such as some of the quantum linear regression and least squares algorithms we discussed in the previous section (e.g.~\cite{rebentrost2014quantum,schuld2016prediction}), the time complexity depends inverse polynomially on the error and the matching procedure has important consequences.
In the next section we discuss these implications and show that, in order to obtain an optimisation error that scales at most as the best estimation error, one should expect to pay a computational price which is polynomial in $n$.

For quantum algorithms with polylogarithmic error dependency, such as~\cite{chakraborty2018power}, the optimisation error is lower than the estimation error and therefore there are no bounds to be matched.
In this case, we show that quantum algorithms argument cannot achieve polylogarithmic runtime in the dimension of the training set based on an argument that analyses the error dependency introduced via the finite sampling process that is required to extract a classical output from the algorithm.
This will be discussed in a later section.


We begin by discussing the dependency on the error and then proceed to discuss the dependency on the measurement errors.
We summarise the results of our analysis in Table~\ref{fig:summary_algos}.

\begin{table}
\begin{center}
    \begin{tabular}{ l | l | l | l  }
     & Algorithm & Train time & Test time \\
    \hline
    Classical & SVM / KRR & $n^3$ & n \\
    & KRR~\cite{yang2017randomized,ma2017diving,gonen2016solving,avron2017faster,fasshauer2012stable} & $n^2$ & $n$ \\
     & Divide and conquer~\cite{zhang2013divide} & $n^2$ & $n$\\
    & Nystr\"om~\cite{williams2001using,rudi2015less} & $n^2$ & $\sqrt{n}$\\
    & FALKON \cite{rudi2017falkon} & $n \sqrt{n}$ & $\sqrt{n}$ \\
     \hline
    Quantum & QKLS / QKLR \cite{chakraborty2018power}& $ \sqrt{n}$ & $ n \sqrt{n}$ \\
    & QSVM \cite{rebentrost2014quantum} & $n \sqrt{n}$ & $n^2 \sqrt{n}$ \\
    \end{tabular}
\end{center}
\caption{Summary of time complexities for training and testing of different classical and quantum algorithms when statistical guarantees are taken into account. We omit $\polylog(n,d)$ dependencies for the quantum algorithms.  
We assume that the generalisation error scales as $ \Theta(1/\sqrt{n})$ and count the effects of measurement errors.
The acronyms in the table refer to: support vector machines (SVM), kernel ridge regression (KRR), quantum kernel least squares (QKLS), quantum kernel linear regression (QKLR), and quantum support vector machines (QSVM).
Note that for quantum algorithms the state obtained after training cannot be maintained or copied and the algorithm must be retrained after each test round. %
This brings a factor proportional to the train time in the test time of quantum algorithms.
%
%
Because the condition number may also depend on $n$ and for quantum algorithms this dependency may be worse, the overall scaling of the quantum algorithms may be slower than the classical.
\label{fig:summary_algos}}
\end{table}

\paragraph{Error dependency of the quantum algorithms}
In this section we show that in order to have a total error (see \cref{eq:totalerrordec}) that scales as $1/\sqrt{n}$ we must introduce a polynomial $n$-dependency in the quantum algorithm.
For simplicity, we present our argument by discussing the case of quantum least squares algorithms with inverse polynomial dependency on the error~\cite{wiebe2012quantum,schuld2016prediction,wang2017quantum}.
Our results generalize easily generalise for all kernel methods.

For a $\gamma$ error guarantee on the final output state, the quantum algorithms we consider have a time complexity that scales as $O(\kappa^c \gamma^{-\beta} \polylog(n))$ for some $\beta, \, c>0$. For example, $\beta=3$ in of \cite{schuld2016prediction} and $\beta=4$ in \cite{li2019sublinear}.

Since for the quantum algorithm the data matrix needs either to be Hermitian or encoded in a larger Hermitian matrix such that the dimensionality of the matrix is $n \times d$ for $n$ data points in $\mathbb{R}^d$,
we assume here for simplicity that the data is given by a $n \times n$ Hermitian matrix, i.e., $n$ points in $\mathbb{R}^n$.

In order give a precise bound to the optimisation error term in \cref{eq:totalerrordec} in terms of the approximation error of the quantum algorithm we consider the following decomposition between the ideal minimizer of the empirical risk $\hat{f}$ and the approximate minimizer $\hat f_\gamma$, output of the learning algorithm
    \begin{align}
\label{eq:algoerror}
    \Ecal( \hat f_\gamma) & - \Ecal( \hat f) \nonumber \\
 &=   \underbrace{\Ecal( \hat f_{\gamma}) - \hat \Ecal( \hat f_{\gamma})}_{\textrm{Generalisation error}} ~+~ \underbrace{\hat \Ecal( \hat f_{\gamma}) - \hat \Ecal( \hat f)}_{\textrm{Algorithmic error}} ~+~ \underbrace{ \hat\Ecal( \hat f) - \Ecal( \hat f)}_{\textrm{Generalisation error}} \\
     &= \Theta(n^{-1/2}) + \underbrace{\hat \Ecal( \hat f_{\gamma}) - \hat \Ecal( \hat f)}_{\textrm{Algorithmic error}},
\end{align}
where the first and third contributions result from the generalisation error bounds and the second is the approximation error of the quantum algorithm. 
In order to achieve the best statistical performance the algorithmic error must scale at worst as the worst statistical error, that is $\hat \Ecal( \hat f_{\gamma}) - \hat \Ecal( \hat f) = O(n^{-1/2})$.

Let us analyse the algorithmic error term for the problem of linear regression and least squares problem.
Assuming that the output of the quantum algorithm is a state $\ket{\tilde{w}}$ while the exact minimizer of the empirical risk is $\ket{w}$, with $\norm{\ket{\tilde{w}} - \ket{w}} \leq \gamma$,
we find that (assuming $|X|$ and $|Y|$ are bounded)   
\begin{align}
    \lvert \hat \Ecal( \hat f_{\epsilon}) - \hat \Ecal( \hat f) \rvert &\leq \frac{1}{n} \sum_{i=1}^n  \left\lvert \left(\tilde{w}^T  x_i -y_i \right)^2 - \left( w^T  x_i -y_i \right)^2 \right\rvert  \\
    & \leq \frac{1}{n} \sum_{i=1}^n  L \left\lvert \left (\tilde{w} - w\right) ^T x_i \right\rvert  \\
    &\leq \frac{1}{n} \sum_{i=1}^n L \norm{\tilde{w} - w} \, \norm{x_i} \leq k \cdot \gamma,
\end{align}
where $k>0$ is a constant and the inequality comes from Cauchy-Schwarz and the fact that, because $|X|$ and $|Y|$ are bounded, we have that, for the square loss $\ell_{\mathrm{sq}}$, the following inequality holds $|\ell_{\mathrm{sq}}(f(x_1), y_1) - \ell_{\mathrm{sq}}(f(x_2),y_2)| \leq L|(f(x_1) - y_1) - (f(x_2) - y_2)|$ for some $L > 0$.

In order to have an algorithm that achieves the best possible statistical accuracy, we need the algorithmic error to scale at worst as the statistical error---this can be obtained by setting $\gamma = n^{-1/2}$.
In this case, the time complexity of quantum least squares becomes
\begin{equation}
    O \left(\kappa^{c} n^{\beta/2} \log(n) \right) ,
\end{equation}
for some constant $c$.

\paragraph{Measurement errors in quantum algorithms}
So far we have ignored the error introduced by the measurement process used to compute a classical estimate of the output of the quantum algorithm.
In practice, this corresponds to the estimation of expected values of quantum operators.
With a classical statistical analysis of the errors---and assuming the measurements are statistically independent---it is possible to show, using the central limit theorem, that the estimation error for a quantum expected value scales as $1/\sqrt{m}$, where $m$ is the number of measurements~\cite{giovannetti2004quantum}. 
This is known as the \textit{standard quantum limit} or the \textit{shot-noise
limit}.
Using techniques developed within the field of quantum metrology it is often possible to overcome this limit---using the same physical resources and the addition of quantum effects such as entanglement---and obtain a precision that scales as $1/m$.
It is possible to show that this is the ultimate limit to measurement precision and follows directly from the Heisenberg uncertainty principle~\cite{giovannetti2004quantum, giovannetti2006quantum}.

In this section we analyse the contribution of the measurement error to the time complexity of quantum learning algorithms.
Let us consider again the case of quantum least squares.
The (quantum) output of the algorithm is the state $\ket{\tilde{w}}$, an approximation (due to algorithmic errors) of the ideal output $\ket{w}$. Using techniques such as quantum state tomography we can produce a classical estimate $\hat{w}$ of the vector $\tilde{w}$ with accuracy
\begin{equation}
\norm{\tilde{w} - \hat{w}} \leq \tau = \Omega(1/m), 
\end{equation}
where $m$ is the number of measurements performed for the estimation of the expected values on $\ket{\tilde{w}}$.

Let $y$ be the ideal prediction. We have two sources of error, the algorithmic error and the error coming from the estimation process
\begin{align}
|y-\hat{y}| &= |w^T x - \hat{w}^T x| \\
& \leq \norm{w-\tilde{w} + \tau} \, \norm{x} \\
& \leq (\gamma + \tau)\, \norm{x}
\end{align}
where we used Cauchy-Schwarz and $\norm{w-\tilde{w}} \leq  \gamma$.

By virtue of \cref{eq:algoerror}, we have that, if we want an algorithm that attains the best statistical accuracy for the number of samples contained in the training set, we need to make sure that the contribution coming from the measurement error scales at most as the worst possible generalisation error. 
Recalling that the generalisation error scales as $\Theta(1/\sqrt{n})$ we have that $\tau=O(1/\sqrt{n})$, from which it follows that $m = \Omega(\sqrt{n})$.
This lower bound on the number of measurement required to extract a classical estimate of the output effectively sets a $\Omega(\sqrt{n})$ lower bound on the time complexity of all supervised quantum machine learning algorithms.

If we consider this lower bound, classical algorithms can have time complexities matching those of the quantum algorithms.
For an more detailed comparison of the runtime of popular classical and quantum algorithms for supervised learning problems see Table~\ref{fig:summary_algos}.

\paragraph{Conclusions}
Quantum machine algorithms promise to be exponentially faster than classical methods. In this paper, we use standard results from statistical learning theory to rule out quantum machine algorithms with polylogarithmic time complexity in the input dimensions.
Considering that almost any current and practically used machine learning algorithm has polynomial runtime, our results warn against the possibility of superpolynomial advantages for supervised quantum machine learning.
We remark two limitations of our analysis. First, our results do not rule out exponential advantages over classical algorithms with superpolynomial runtime. 
Second, we do not make assumptions on the hypothesis space; using prior knowledge it is possible get error rates that converge faster than $1/\sqrt{n}$. 

Our argument leverages the fact that the statistical error of the algorithm has a provable 
polynomial dependence on the number of samples in the training set. Since the statistical error and the 
approximation error of the algorithm are additive, in order to 
achieve the best possible error rate, the asymptotic scaling of the statistical error
must match that of the approximation error. This matching forces the approximation error of quantum
algorithms to scale polynomially with the number of samples. This effectively kills
quantum speedups for algorithms that have polynomial dependence on the error.

For algorithms where the dependency on the error is logarithmic, this argument does not
apply. In this case, we show that the sampling error coming from the measurement process also
adds up additively to the total error and this introduces
a polynomial dependency in the number of samples that kills the superpolynomial speedup.

Notably, our results hold even assuming that quantum algorithms can access a quantum data
structure at no cost. In this respect, we prove a stronger `no-go' result for quantum learning
than the one proved by Tang in~\cite{tang2018quantum}.
Indeed, the latter relies on a classical data structure that
mimics a quantum data structure but is unrealistic in practice.    

As future directions, it is worth mentioning that it may be possible strengthen our results by analysing the $n$ dependency of the condition number. Previous results in this direction are discussed in~\cite{cucker2002mathematical,hochstadt2011integral}.

The authors would like to thank Aram Harrow, Sathya Subramanian, and Maria Schuld for helpful feedback on an earlier draft of the article, Shantanav Chakraborty and Stacey Jeffrey for discussions on the Frobenius norm dependency of their quantum least squares algorithm, and Alessandro Davide Ialongo for helpful comments.
This research was supported in part by the National Science
Foundation under Grant No. NSF PHY-1748958 and by the Heising-Simons
Foundation.
A.R. is supported by the Simons Foundation through \textit{It from Qubit: Simons Collaboration on Quantum Fields, Gravity, and Information}.
L.W.\ is supported by a Google PhD Fellowship.

\bibliography{quantum_learning}

\end{document}